# Multi-State Trajectory Approach to Non-Adiabatic Dynamics: General Formalism and the Active State Trajectory Approximation

*Guohua Tao**

Shenzhen Key Laboratory of New Energy Materials by Design, Peking University, Shenzhen, China 518055

School of Advanced Materials, Peking University Shenzhen Graduate School, Shenzhen, China 518055

* Corresponding author: taogh@pkusz.edu.cn

# ABSTRACT

A general theoretical framework is derived for the recently developed multi-state trajectory (MST) approach from the time dependent Schrödinger equation, resulting in equations of motion for coupled nuclear-electronic dynamics equivalent to Hamilton dynamics or Heisenberg equation based on a new multistate Meyer-Miller (MM) model. The derived MST formalism incorporates both diabatic and adiabatic representations as limiting cases, and reduces to Ehrenfest or Born-Oppenheimer dynamics in the mean field or the single state limits, respectively. By quantizing nuclear dynamics to a particular active state, the MST algorithm does not suffer from the instability caused by the negative instant electronic population variables unlike the standard MM dynamics. Furthermore the multistate representation for electron coupled nuclear dynamics with each state associated with one individual trajectory presumably captures single state dynamics better than the mean field description. The coupled electronic-nuclear coherence is incorporated consistently in the MST framework with no ad-hoc state switch and the associated momentum adjustment or parameters for artificial decoherence, unlike the original or modified surface hopping treatments. The implementation of the MST approach to benchmark problems shows reasonably good agreement with exact quantum calculations, and the results in both representations are similar in accuracy. The active state trajectory (AST) approximation of the MST approach provides a consistent interpretation to trajectory surface hopping, which predicts the transition probabilities reasonably well for multiple nonadiabatic transitions and conical intersection problems.

I. Introduction

Nonadiabatic dynamics involving transitions among multiple states is crucial to many key molecular processes in the fields of chemistry, biology and materials. A variety of theoretical methods[1-4] have been developed to describe nonadiabatic dynamics at the molecular level aiming to provide physical insights into related processes, among which mixed quantum-classical methods such as Ehrenfest dynamics,[5-11] surface hopping,[12-27], extended Born-Oppenheimer (BO) dynamics,[28-31] quantum-classical Liouville dynamics,[32-36] wave-packet-based methods,[37-44] quantum trajectory methods,[45-48] semiclassical approaches,[49-60] path-integral-derived methods,[61-63] and recently developed symmetrical quasiclassical (SQC) Meyer-Miller (MM) approaches,[64-71] to name a few, have great potential in practical implementations. However, performing efficient and accurate nonadiabatic dynamics simulations of large molecular systems is still a challenge.

One requirement to achieve the desired accuracy in nonadiabatic molecular dynamics simulations is to treat the coupled electronic-nuclear coherence properly. Ehrenfest dynamics overestimates electronic coherence and nuclear trajectories always evolve on an average potential energy surface of multiple states. Surface hopping (in particular in Tully's fewest switches formulation[12]) propagates nuclear dynamics on a single state except for a stochastically local state switch based on a hopping criterion therefore avoids the incorrectness of the mean field description on nuclear motions in the asymptotic regime. Surface hopping is certainly very useful and popular due to its simplicity. However it also suffers from the problem of overcoherence, not because of running electronic dynamics coherently, but due to the overestimation of nuclear coherence since the interstate couplings are evaluated by the single state trajectory locally and the coupled electronic dynamics is propagated incorrectly (see discussions below). Extensive studies[9,16-21] have been devoted to introduce decoherence to either electronic dynamics or nuclear motions, by adding a decoherence term to account for the electronic-nuclear couplings,[9] or interstate nuclear coherence related to the Frank-Condon factor[16-20,51] or nuclear phase.[21] Most of these efforts focus on the determination of a phenomenological decay time for the decoherence term.

Developed nearly forty years ago, the original MM approach[50] maps electronic degrees of freedom (DOFs) into classical variables resulting in a consistent treatment on both electronic and nuclear DOFs at the same dynamical footing thus has the advantage of properly describing electronic-nuclear coherence in nonadiabatic dynamics. In combination with the semiclassical (SC) initial value representation (IVR),[72] the MM approach has been implemented to a variety of nonadiabatic processes[57-62] with a great success. However the computational cost of the SC-IVR treatment is still expensive for large systems if no further approximation is made. For example, it has been shown[73] that by means of an efficient correlated importance sampling scheme, the SC-IVR/MM method is capable of describing a model photosynthesis system consisting of two states and ten bath modes at the full semiclassical level. Very recently Miller and coworkers proposed the SQC/MM approach[64-67] to allow nonadiabatic dynamics simulations of complex molecular systems[68-71] to be performed at the quasiclassical level as efficient as classical molecular dynamics simulations. One major problem with the MM model and related nonadiabatic simulation methods is the mean field description on nuclear dynamics, i.e. nuclear motions are governed by an averaged force of multiple potential energy surfaces weighted by the corresponding electronic population variables. Since the instant electronic population variable of an individual trajectory may be negative, the MM trajectory could be unstable and the algorithm becomes numerically inefficient.

In the hope of finding a better way to deal with realistic molecular systems, recently we proposed a multi-state trajectory (MST) approach[74] to nonadiabatic dynamics simulations. In the MST representation, each state is associated with one individual nuclear trajectory no matter if the state is active (i.e. the system is on this state) or not, and nuclear dynamics is governed mainly by the corresponding potential energy surface (plus the interstate couplings or nonadiabatic couplings). Therefore the coupled electronic-nuclear dynamics could be presumably better represented in terms of the state-specific nuclear motions. Furthermore, by quantizing nuclear dynamics to the active state using a windowing technology proposed in the SQC/MM approach,[65] the MST method avoids counting the contribution from the state with an instant negative electronic population variable resulting in a very stable working algorithm for nonadiabatic molecular dynamics simulations on complex systems especially for those involving many electronic states,[74,75] unlike the mean-field treatment on the nuclear forces in the standard SQC approach. When the quasi-classical (QC) approximation is made, i.e. nuclear dynamics is approximated by an ensemble of classical trajectories instead of wavepackets, it has been demonstrated that the resulting MST-QC treatment can describe nonadiabatic dynamics of complex molecular systems quite efficiently, ex. nonequilibrium nonadiabatic molecular dynamics in a model system with 12 states and 671 atoms for electron coupled lithium ion diffusion in cathode materials $LiFePO_4$ for batteries was simulated for a time on the order of ten picoseconds.[75]

In previous work,[74] the MST-QC approach was applied to several benchmark model systems such as the spin-boson problem, and single/dual avoided crossing problems in Tully's original paper[12] on fewest switches surface hopping (FSSH), and the nonadiabatic dynamics simulations were performed only in the diabatic representation. However sometimes it is more convenient to consider the adiabatic representation. In fact, standard ab initio electronic structure calculations make it available the state energies and nonadiabatic couplings associated with adiabatic wave functions, which in principle could be used for ab initio nonadiabatic dynamics simulations straightforwardly. In this work, we provide the general theoretical framework of the MST approach incorporating both diabatic and adiabatic representations, based on which further developments or approximations could be made therefore the applicability of the method would be extended. For example, when the only active state trajectory (AST) ensemble is considered, an augmented image (AI) version of MST may be achieved, which could better account for the multiple transitions in open channel problems such as Tully's extended coupling model. Furthermore, the AST (or MST-AI) with the coupling terms evaluated locally switches equivalently one single trajectory among different states therefore may provide an alternative interpretation to surface hopping.

Because surface hopping especially FSSH is so popular that continuous efforts[13,24,47] have been made on the derivation of surface hopping algorithm. One conventional way[3,13,47,48] to achieve classical or classical-like dynamics from a full quantum description, say Schrödinger equation, is to apply Bohm's formulation[76] using a polar representation of nuclear wave function. The resulting Bohmian dynamics features a quantum potential, which is suggested to be a key factor in reproducing the full nuclear quantum effect. Following these pioneer work,[3,13,47,48] we also adopt the Bohm's formulation in the derivation of our (electronic coupled) nuclear dynamics. However, our strategy is different. In the MST representation, the nuclear trajectory associated with the corresponding state equivalently defines an adiabatic nuclear basis, and the time evolution of the system is performed by the unitary transformation of the system Hamiltonian. Such a Heisenberg picture results in nuclear equations of motion (EOM) free of nuclear state mixing[3,13,40,47,48] and the consequent numerical instability,[40,47] unlike that in the conventional treatment. Note that the adiabatic nuclear basis here we used is different from the electronically adiabatic representation, so that our dynamics is fully represented by a set of electronic-nuclear coupled equations, unlike the extended BO dynamics.[29,30]

There are quite a few multiple trajectory methods[20,21,40-44,77] to be implemented to a variety of models systems successfully. For example, Shenvi, Subtonik, and Yang proposed the phase corrected surface hopping approach[21] and Adhikari and coworkers developed the time-dependent discrete variable representation (TDDVR),[42,43] and both produce results in excellent agreement with exact calculations on Tully's models. However previous work on surface hopping with multiple trajectories either focuses on the evaluation of the decoherence time instead of dynamics,[20] or introduces an intuitively derived momentum dependent Hamiltonian.[21] TDDVR normally requires a number of grid points (multi-trajectories) for each mode to achieve accurate results, which may become inefficient if the wave function is delocalized or the particles dynamics is delocalized or the many body interactions are important especially in complex systems so that many grid points may be needed. Moreover both TDDVR and multiconfigurational Ehrenfest approach[44] propagate Ehrenfest dynamics. With many successful implementations, multi-spawning methods[40,41] due to Martinez and coworkers and path branching methods[77] developed by Takatsuka and coworkers feature in propagating multiple trajectories, however care has to be taken in setting the criteria of generating new trajectories, and the number of trajectories on a single surface may grow rapidly so that the efficiency of these methods for complex systems may be limited. While the MST approach is distinct from these methods on that not only it can be derived rigorously from the time-dependent Schrödinger equation in a general framework with no adjustable parameter but it may have advantages of retaining its efficiency and numerical stability in simulating complex molecular systems. Even though the standard quasiclassical version of MST may not achieve high accuracy in describing nuclear quantum effects, associating more trajectories with each state (a multiconfiguration description) or using semiclassical trajectory may improve the accuracy.

Section II derives the MST formalism in a general framework starting from the time dependent Schrödinger equation. Models of Tully's nonadiabatic problems,[12] the quasi-Jahn-Teller problem[28] and a spin-boson problem[23] are presented in Sec. III along with computational details. Sec. IV provides results and discussions on the implementation of the MST and MST-AI to benchmark problems, which is summarized in Sec. V.

**II.  Theory**

We start from writing the total nuclear-electronic wavefunction in a multistate representation, i.e.

$$\Psi(r, R, t) = \sum_{k=1}^{F} a_k(t)\phi_k(r, R)\chi_k(R, t), \quad (1)$$

where $\Psi$ is the total wavefunction, $a_k, \phi_k, \chi_k$ are the complex coefficient, the normalized electronic wavefunction, and the normalized nuclear wavefunction for the state k, respectively, and F is the number of the state of the system. Plug Eq. 1 into the time dependent Schrödinger equation,

$$i\hbar \frac{\partial \Psi}{\partial t} = \mathcal{H}\Psi, \quad (2a)$$

we have

$$i\hbar \sum_{k=1}^{F} \dot{a}_k(t)\phi_k(r, R)\chi_k(R, t) + a_k(t)\phi_k(r, R)\dot{\chi}_k(R, t) =$$

$$\mathcal{H} \sum_{k=1}^{F} a_k(t)\phi_k(r, R)\chi_k(R, t) \ , \quad (2b)$$

with $\mathcal{H}$ the system Hamiltonian, $\hbar$ the Planck constant over $2\pi$, and the dot represents for the time derivative. Within the time dependent self-consistent field (TDSCF) formalism,[52,78] the EOM for the coupled nuclear-electronic system read

$$i\hbar \dot{a}_k(t) = \sum_{l=1}^{F} a_l(t)\langle \chi_k | \mathcal{H}_{kl}^r | \chi_l \rangle \ , \tag{3a}$$

$$i\hbar \dot{\chi}_k = \mathcal{H}\chi_k \ , \tag{3b}$$

here we denote $\chi_k = \chi_k(R,t)$ and $\mathcal{H}_{kl}^r = \langle \phi_k(r,R) | \mathcal{H} | \phi_l(r,R) \rangle_r$. The equations of motion for the nuclear degrees of freedom (DOF) Eq 3b can be written explicitly

$$i\hbar \dot{\chi}_k = \sum_{n,m}^{F} \mathcal{H}_{mn}\chi_k = \sum_{m=1}^{F} \mathcal{H}_{mk}\chi_k \ , \tag{4}$$

with the Hamiltonian matrix element

$$\mathcal{H}_{mk} = T_{mk}^R + |m\rangle\langle k| \left( T_{mk}^{r(2)} + T_{mk}^{r(1)} + V_{mk}^r \right), T_{mk}^R = -\frac{\hbar^2}{2\mu}\nabla_k^2 \delta_{mk} \ , T_{mk}^{r(2)} =$$

$$-\frac{\hbar^2}{2\mu}\langle \phi_m | \nabla^2 | \phi_k \rangle = -\frac{\hbar^2}{2\mu} D_{mk} \ , T_{mk}^{r(1)} = -\frac{\hbar^2}{\mu}\langle \phi_m | \nabla | \phi_k \rangle \nabla = -\frac{\hbar^2}{\mu} d_{mk}\nabla, \text{and } V_{mk}^r =$$

$$\langle \phi_m | V | \phi_k \rangle = V_{mk}. \tag{5}$$

Here $\mu$ is the reduced nuclear mass, and $d_{mk}$, $D_{mk}$ are the first and second order nonadiabatic coupling terms.[2] Eq. 3 is a set of electronic-nuclear coupled EOM, and interstate transfer is governed by the off-diagonal Hamiltonian matrix elements. Nuclear dynamics Eq. 3b (and Eq. 4) is represented in terms of adiabatic nuclear basis on which are projected all contributions of interstate couplings. This is in contrast to conventional TDSCF formalisms,[13] in which every (time-dependent) nuclear basis is projected onto the whole basis set.

From Eq. 3a, the EOM for electronic complex coefficients matrix may be obtained

$$i\hbar \dot{B}_{mk} = \sum_{l=1}^{F} [B_{ml}\langle \chi_l | \mathcal{H}_{lk}^r | \chi_k \rangle - \langle \chi_m | \mathcal{H}_{ml}^r | \chi_l \rangle B_{lk}] \ , \tag{6}$$

with $B_{mk} \equiv a_m^* a_k$, which unifies the population, coherence and flux functions, i.e. $h_k = c_{kk} = B_{kk}, c_{mk} = \text{Re}[B_{mk}], f_{mk} = \text{Im}[B_{mk}]$. And the nuclear EOM may be written by (see the supplement[79] for the derivation)

$$\frac{\partial P_k}{\partial t} = -\nabla H_k, H_k = \frac{1}{2\mu}P_k^2 + \sum_m \left[c_{mk}V_{mk} + f_{mk}\frac{\hbar}{\mu}d_{mk}P_k\right], \text{ and} \tag{7a}$$

$$\frac{\partial Q_k}{\partial t} = \frac{J_k}{\rho_k} = \frac{P_k}{\mu} + \sum_m \frac{\hbar}{\mu}f_{mk}d_{mk}. \tag{7b}$$

Eq. 6, and Eq. 7 are the EOM for the coupled nuclear-electronic system from the time dependent Schrödinger equation, which may also be derived from the Hamilton dynamics based on the MM model.[50,74] According to Eq. 7, we propose the multistate MM Hamiltonian as the follows:

$$H = \sum_{k=1}^{F} H_k = \sum_{k=1}^{F} \frac{1}{2\mu}P_k^2 + \sum_{m,k} \text{Re}\left[B_{mk}\left(V_{mk} - \frac{i\hbar}{\mu}d_{mk}P_k\right)\right], \tag{8a}$$

where the electronic coefficient matrix

$$\hat{B}_{mk} = a_m^\dagger a_k = \frac{1}{2}[(\hat{x}_m\hat{x}_k + \hat{p}_m\hat{p}_k) + i(\hat{x}_m\hat{p}_k - \hat{p}_m\hat{x}_k)], \tag{8b}$$

and $a_m^\dagger = \frac{1}{\sqrt{2}}(\hat{x}_m - i\hat{p}_m), a_m = \frac{1}{\sqrt{2}}(\hat{x}_m + i\hat{p}_m).$ (8c)

The EOM for the coupled nuclear-electronic dynamics may alternatively be obtained from Heisenberg equation[27,80]

$$\frac{dB}{dt} = \frac{1}{i\hbar}[B, H], \tag{9a}$$

or $i\hbar\dot{B}_{mk} = \sum_{l=1}^{F}[B_{ml}\mathcal{H}_{lk}^{r,R} - \mathcal{H}_{ml}^{r,R}B_{lk}],$ (9b)

with $\mathcal{H}_{mk}^{r,R} = \langle\chi_m|\mathcal{H}_{mk}^r|\chi_k\rangle$;

$$\frac{dQ_k}{dt} = \frac{1}{i\hbar}[Q_k, H] = \frac{\partial H}{\partial P_k}, \tag{9c}$$

$$\frac{dP_k}{dt} = \frac{1}{i\hbar}[P_k, H] = -\frac{\partial H}{\partial Q_k}. \tag{9d}$$

It is interesting to see that the mean field Ehrenfest description[5] and Born-Oppenheimer (BO) dynamics[81] can be obtained as limiting cases[74] from the above derivation for the multistate trajectory approach. Indeed if there is only one individual trajectory, i.e.

$\chi_k(R, t) = \chi(R, t)$ for all k, (10a)

then $\quad \Psi(r, R, t) = \sum_{k=1}^{F} a_k(t)\phi_k(r, R)\chi(R, t)$ , (10b)

$H = \frac{1}{2\mu}P^2 + \sum_{m,k} \text{Re}\left[B_{mk}\left(V_{mk} - \frac{i\hbar}{\mu}d_{mk}P\right)\right]$ , (10c)

$i\hbar\dot{B}_{mk} = \sum_{l=1}^{F}[B_{ml}\mathcal{H}_{lk}^r - \mathcal{H}_{ml}^r B_{lk}]$ , (10d)

$\frac{\partial P}{\partial t} = -\nabla H$ , (10e)

$\frac{\partial Q}{\partial t} = -\frac{\partial H}{\partial P}$ . (10f)

Here we have the mean field wave function Eq. 10b, the standard MM Hamiltonian Eq. 10c, the corresponding EOM for electronic DOF Eq. 10d, for nuclear DOF (Ehrenfest dynamics) Eq. 10e, and Eq. 10f. Instead if the system evolves on a single state only, say the state k, then

$\chi_m(R, t) = 0$, and $\mathcal{H}_{mk} = 0$ for all $m \neq k$, (11a)

and the total wave function now includes only a single term in Eq. 1,

$\Psi(r, R, t) = a_k(t)\phi_k(r, R)\chi_k(R, t)$ , (11b)

$H = H_k = \frac{1}{2\mu}P_k^2 + V_{kk}$ , (11c)

$\frac{\partial P_k}{\partial t} = -\nabla H_k$ , (11d)

$\frac{\partial Q_k}{\partial t} = -\frac{\partial H_k}{\partial P_k}$ . (11e)

Eqs. 11c-e are the Hamiltonian, and EOM for BO dynamics.

For coupled nuclear-electronic dynamics, it is essential to describe coherence properly.[82] Coherence comes from the interference between different states, including electronic states, vibrational states, rotational states, or translational states etc. Here we focus on electronic state, and group all nuclear motions into intra-state dynamics. Correspondingly coherence may be classified into electronic coherence and nuclear coherence, the latter of which includes intra-state nuclear coherence, and inter-state nuclear coherence. Electronic coherence and nuclear coherence may be coupled to each other, ex. vibronic coherence, which may be identified by ultrafast spectroscopy technology, as demonstrated by Shi and coworkers.[83] Ehrenfest dynamics proceeds on an average potential energy surface all the time, so that different electronic states are coupled through nuclear motions even when the interstate couplings diminish, which effectively exaggerates electronic coherence (Eq. 10). No inter-state nuclear coherence exists in the Ehrenfest description since there is only one nuclear trajectory. On the other hand, BO dynamics evolves on a single state and completely ignores both electronic coherence and inter-state nuclear coherence (Eq. 11). Surface hopping propagates nuclear dynamics on a particular state in a BO style until an inter-state hopping takes place, while the electronic dynamics does not contribute explicitly to nuclear EOM except for providing an electronic probability based hopping criterion from electronic EOM similar to Eq. 3a or Eq. 6. The difference is that in surface hopping the off-diagonal Hamiltonian matrix elements $\langle \chi_k | \mathcal{H}_{kl}^r | \chi_l \rangle$ are evaluated locally from the same one individual nuclear trajectory, which effectively overestimates inter-state nuclear coherence and destroy the coupled nuclear-electronic coherence. Introducing decoherence due to the Frank-Condon effect[16,17,51] similar to the damping function Eq. S8c alleviates the overestimation, while Eq. S8a in the MST approach and similar treatment in the multiple spawning method[40] apparently could make a better approximation to the off-diagonal Hamiltonian matrix elements, which in turn results in a better description on electronic coherence. In principle, both intra-state and inter-state nuclear coherence may be captured well by averaging an ensemble of semiclassical trajectories, as normally adopted in the SCIVR methods.[72] Under the assumption that the multistate trajectory is purely classical with no nuclear phase, Eq. S8a may still provide a good approximation comparably with the localized classical description in surface hopping.

Once the mixed quantum-classical or quasi-classical approximations are applied, the accuracy of description of inter-state couplings and nuclear dynamics may depends on which representation is used, either adiabatic or diabatic.[4] In the diabatic representation, the inter-state couplings (off-diagonal Hamiltonian matrix elements) $V_{mk}$ are inherently nonlocal; by contrast in the adiabatic representation, the inter-state couplings $-\frac{i\hbar}{\mu}d_{mk}P_k$ contain local gradient operators. Therefore presumably a local classical treatment on nuclear dynamics would be more accurate in the adiabatic representation than that in the diabatic representation, which is consistent with the experience on the implementation of surface hopping.[4] For the quasiclassical approximation of the MST approach, we expect a similar trend (see below). In the standard version of MST, we adopt the following treatments:

$$\dot{P}_k = -\nabla \sum_m \left[\frac{1}{\mu}c_{mk}V_{mk}(Q_k) + f_{mk}\frac{\hbar}{\mu}d_{mk}(Q_k)P_k\right] , \qquad (12a)$$

$$\mathcal{H}_{mk}^{r,R} = \frac{1}{2}\left(-\frac{i\hbar}{\mu}d_{mk}(Q_m)P_m - \frac{i\hbar}{\mu}d_{mk}(Q_k)P_k\right) , \qquad (12b)$$

in the adiabatic representation, and

$$\dot{P}_k = -\nabla \sum_m \left[\frac{1}{\mu}c_{mk}V_{mk}(\overline{Q}_{mk})\right] , \qquad (13a)$$

$$\mathcal{H}_{mk}^{r,R} = V_{mk}(\overline{Q}_{mk}) , \qquad (13b)$$

in the diabatic representation. Here we make it sure the system Hamiltonian is Hermitian. This construction assures the microscopic reversibility holds.

To properly describe the wave function bifurcation upon nonadiabatic transitions, direct treatments were proposed such as wave function spawning,[20,40] and path branching.[3] Alternatively the MST representation suggests a statistical average picture based on the multistate trajectory ensemble, i.e. a collection of state specific trajectories which never bifurcate, and the quantum wave function bifurcation is represented by the ensemble average of the active state trajectories. However there may have a problem with the current version of the MST approach, that is individual state-specific nuclear trajectory could move apart on different potential energy surfaces but never come back, for example in the unbounded scattering problems such as Tully's model, therefore multiple nonadiabatic transitions may not be correctly accounted for. In such cases, MST may not describe nuclear coherence well and tends to underestimate the local contribution of nuclear dynamics on nonadiabatic transitions. Ideally if multiple trajectories are allowed to be associated with each state (a multiconfiguration MST) so that there are neighboring trajectories available on other states around the transition regime, the inter-state couplings and therefore the coupled electronic-nuclear dynamics could be better described. However this idea (in the same spirit to the one in multiple spawning) would require a number of configurations and may make the algorithm for multiple transitions in high dimensional systems inefficient.

To account for the locality of nonadiabatic transitions while keep the trajectory picture as simple as possible, we approximate the Hamiltonian matrix elements by the nuclear trajectory on the active state k and its augmented image (AI) in the coupled state m ≠ k, i.e.

$$\langle \chi_m | \mathcal{H}_{mk}^r | \chi_k \rangle = \langle \chi_m(Q_m) | \mathcal{H}_{mk}^r | \chi_k(Q_k) \rangle \, , \tag{14a}$$

and $Q_m = Q_k, P_m = P_k$ for all $m \neq k$ . (14b)

When a transition occurs from the state k to the state m, the trajectory on the state k is deactivated and the state m becomes the new active state. In comparison with the standard MST approach, the augmented image in the MST-AI treatment is not propagated on the inactive state but just used for the evaluation of the nonadiabatic couplings. Effectively the MST-AI treatment is based only on the active state trajectory ensemble, so that nonadiabatic couplings could be better represented in cases of that trajectories on different states diverge before nonadiabatic transitions take place, at the cost of overestimation of nuclear coherence.

The MST-AI/AST representation of nonadiabatic dynamics indeed resembles the popular surface hopping picture in that at any given time, only one (active) state is occupied and the off-diagonal Hamiltonian matrix elements are evaluated locally based only on the active state trajectory [note that this is different from the more general MST treatment, which requires (nonlocal) multiple state information]; furthermore the trajectory may switch (hop) from one active state (surface) to another upon electronic transitions. However the distinction between the two methods is clear. The AST dynamics is fully electronic-nuclear coupled whereas in surface hopping nuclear dynamics (being purely classical) is decoupled from electronic dynamics.

Moreover the state switch criteria in AST (and MST) is based on the quantization of instantaneous electronic population using the windowing technology rather than the interstate couplings in surface hopping, the latter of which is not a perfect measure for hopping since sometimes the hop may be unsuccessful therefore further ad hoc treatment has to be introduced to account for the true transition probability. By contrast, every success switch (reading from the window function) is guaranteed by the coupled electronic-nuclear dynamics so that no extra momentum adjustment is required, unlike that in surface hopping. In another word, the AST picture provides an illustrative understanding of surface hopping, which we believe a better way than constructing a theoretical derivation for the original ad hoc hopping criteria.

Although our derivation for the MST approach starts from time dependent Schrödinger equation, the idea is closely related to the MM model and its implementation to realistic molecular systems. Indeed electronic dynamics and the quantization of nuclear dynamics are essentially the same as in the standard MM framework[50] and the related SQC approach.[65] The major difference between the MST approach and other methods based on the MM model is the way to treat nuclear dynamics and the MST trajectory is state-specific instead of Ehrenfest,[74] comparing Eq. 7a and Eq. 10e. It is noted that the classical mapping variable of the state occupying number $h_k = B_{kk}$ could be negative (supposed to be 1 or 0 in the quantum description), which may causes serious problems for MM nuclear dynamics. For example, the effective potential of the associated state would be inverted for a negative $h_k$ and the mean filed nuclear dynamics may be dominated by this component and diverges quickly. However, interpreting $h_k$ of an individual MM trajectory directly to the state population is misleading since they are equal only on the ensemble average level. It is more appropriate to associate the individual $\{h_k\}$ with the corresponding domain in the state space,[84] which maps exactly to the quantum state. Therefore the real serious problem with the MM dynamics is the mean field description. By contrast, in the MST treatment,[74] individual trajectories are propagated under the force mainly determined by the associated single state (plus inter-state couplings), and a windowing function is applied (same as in the SQC approach[64]) to map the system to a particular state. The corresponding $h_k$ for the active state is always positive, and only the active state contributes to the ensemble averaged nuclear dynamics. Consequently the MST dynamics is stable and has been successfully implemented to complex molecular systems.[75]

The total energy of the coupled nuclear-electronic system is conserved since the multistate Hamiltonian does not depends on time explicitly, i.e.

$$\frac{dH}{dt} = \frac{\partial H}{\partial t} + \sum_{k=1}^{F} \left( \frac{\partial H}{\partial P_k} \frac{\partial P_k}{\partial t} + \frac{\partial H}{\partial Q_k} \frac{\partial Q_k}{\partial t} \right) = 0 \ . \qquad (15)$$

Comparing with the standard MM Hamiltonian Eq. 10c, the MST Hamiltonian Eq. 8a contains kinetic energy terms for each individual state. However, in the MST approach, only one state is active even though multiple trajectories are propagated simultaneously, and the total (kinetic) energy is actually obtained from an ensemble average of such active state in multistate trajectories. For example, assuming the final distribution $h_k$ is achieved in the asymptotic region, the total energy is given by

$$E = \sum_{k=1}^{F} h_k \left( \frac{1}{2\mu} P_k^2 + V_{kk} \right) \ . \qquad (16)$$

This is similar to the SQC approach, in which proper electronic populations could be recovered by quantizing the Ehrenfest trajectory on the active state only.

### III. Computational details

The MST and its variants are applied in both diabatic and adiabatic representations to Tully's three two-state models.[12] The corresponding potentials in both diabatic and adiabatic representations are shown in Figure S1. The nuclear mass is 2000 a.u., and the initial wave function is a coherent state located on the lower state (state 1), i.e. $\psi(P_0, Q_0) = e^{-(Q-Q_0)^2/\sigma^2 + iP_0(Q-Q_0)}$ with the initial position $Q_0 = -20$ a.u. and the width $\sigma = 20/P_0$.[12] The exact quantum calculations for the benchmark model system were performed on a grid of 2000-9000 points with a spacing of 0.032 a.u. by using the discrete variable representation (DVR) method.[85] For the MST simulations, normally 9600 trajectories are enough to achieve well converged results in both the standard MST and MST-AI treatments although the convergence of the latter is not as good as that of the former. The time step is 5 a.u. and 1 a.u. in the standard MST and MST-AI simulations, respectively, which satisfies the convergence check.

Another common type of interesting nonadiabtic problems involves conical intersection in which the interstate couplings vanish in the regime where different states are degenerate therefore nonadiabatic couplings diverges. Here we consider a quasi-Jahn-Teller model[28,30,43] of scattering processes for a two state and two dimensional system. The two adiabatic potential energy surfaces are:

$$U_1 = \frac{1}{2}\mu[\omega_0 - \omega_1 e^{-(R/\sigma_1)^2}]r^2 + Ae^{-\frac{R^2+r^2}{\sigma^2}};$$

$$U_2 = \frac{1}{2}\mu\omega_0^2 r^2 - (D - A)e^{-\frac{R^2+r^2}{\sigma^2}} + D. \qquad (17a)$$

Here R and r are coordinates representing the reactive and internal DOF, respectively. They can be written in terms of polar coordinates as the follows:

$$R = \rho\cos\theta; \; r = \rho\sin\theta. \qquad (17b)$$

The corresponding diabatic states are defined as:

$$W_{11} = \frac{1}{2}[U_1 + U_2 + (U_1 - U_2)\cos\theta] \; ; \quad W_{22} = \frac{1}{2}[U_1 + U_2 - (U_1 - U_2)\cos\theta] \; ;$$

$$W_{12} = \frac{1}{2}(U_1 - U_2)\sin\theta \;. \qquad (17c)$$

The model parameters[28,43] are $\mu = 0.58$ a.m.u., $\omega_0 = 39.14 \times 10^{13} s^{-1}$, $\omega_1 = 7.83 \times 10^{13} s^{-1}$, $\sigma_1 = 0.75$Å, $\sigma = 0.30$Å, $A = 3.0$ eV, $D = 5.0$ eV.

The third model we consider in this work is a spin-boson problem[23] of an explicit two-state nuclear mode coupled with harmonic bath. The Hamiltonian is given by

$$H = \frac{P^2}{2M} + \frac{1}{2}M\Omega^2 x^2 + \left(\lambda x + \frac{\varepsilon_0}{2}\right)\sigma_z + V\sigma_x + \sum_i \left[\frac{p_i^2}{2m_i} + \frac{1}{2}m_i\omega_i^2\left(z_i - \frac{\zeta_i}{m_i\omega_i^2}x\right)^2\right]. \qquad (18)$$

Here p, x are the nuclear momentum and coordinate, $p_i$ and $z_i$ are the momentum and coordinate of the *i*-th bath mode, and $\sigma_z, \sigma_x$ are Pauli matrices for the two diabatic states. The parameters are taken from ref 23, with mass M=1, frequency $\Omega = 0.00004375$, the diabatic coupling $V = 0.00002$, the reorganization energy $E_r = \frac{2\lambda}{M\Omega^2} = 0.0239$, the driving force $\varepsilon_0 = 0.0239$, damping parameter $\gamma = \frac{\pi\zeta_i^2\rho(\omega_i)}{2m_i\omega_i^2} = 0.00015$, temperature $k_B T = 0.00095$. Here the pure Ohmic spectra density is applied with the density of state $\rho(\omega_i)$.

## IV. Results and Discussion

The standard MST approach and its variants are tested in both diabatic and adiabatic representations on Tully's models. The calculated transmission probabilities for the single and dual avoided crossing problems are shown in Figures S2 and S3, respectively. In both cases, the standard MST approach provides results in reasonably good agreement with the exact quantum calculations with accuracy at a similar level in both diabatic and adiabatic representations. For the single avoided crossing model, the standard MST works best in both representations. Making the QC approximation to nuclear dynamics (Eq. S11) simplifies the algorithm while shows only a little degrade in accuracy in the low momentum regime. Also assuming that interstate couplings in nuclear dynamics are local (Eq. S12) produces similar results with the standard calculations except for the generally small deviations in the low momentum regime. Similar trend is seen for the AI treatment. For the dual avoided crossing problem, the standard MST approach is the best in the diabatic representation and the MST-AI works similarly well, while the QC approximation predicts the transmission probabilities quite poorly in the intermediate energy regime. The local approximation on nuclear dynamics seems to make no difference. In the adiabatic representation, all MST treatments are in excellent agreement with the exact quantum mechanical method. Overall MST-AI performs slightly better than the standard MST, which generates results in similar accuracy as the QC approximation. The local approximation seems detrimental to the prediction in the low energy regime.

The transmission and reflection probabilities were calculated for the extended coupling problem, and the MST results are shown in Figure 1, in comparison with exact quantum DVR calculations. All types of MST treatments predict the transmission probability of the lower state remarkably well, and the transmission probability of the upper state obtained from the QC version of MST is also in excellent agreement with exact quantum calculations (Fig. 1a). Therefore the sum of reflection probabilities of two channels is also well reproduced by the MST-QC. However, MST-QC does not allow for the second nonadiabatic transition due to the diverged nuclear trajectories on different states, resulting in a vanished ground state reflection i.e. $R_{1 \leftarrow 1} = 0$, and an overestimated reflection probability of the excited state (Fig. 1b). The full nuclear dynamics Eq. S9 does not predict correctly the upper state transmission probability $T_{2 \leftarrow 1}$, and the total reflection probability in the high momentum regime. And the effect of local approximation is negligible.

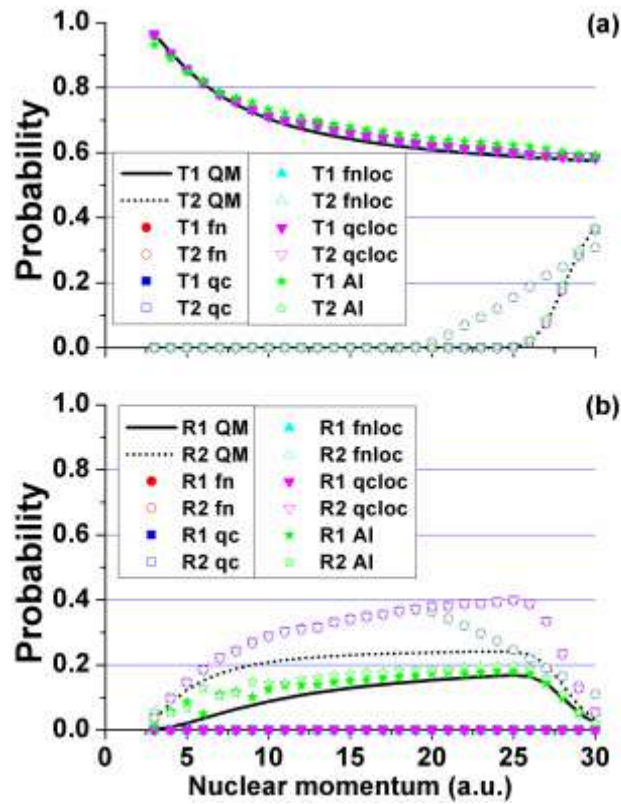

Figure 1. Extended coupling problem. Transmission probability for state 1 (T1) and state 2 (T2) obtained from various MST treatments (full nuclear dynamics Eq. S9, fn, red circles; quasiclassical approximations Eq. S11, qc, blue squares; their combination with local approximation: Eq. S9 with Eq. S12, fnloc, cyan triangle; Eq. S11 with Eq. S12, qcloc, pink inverted triangle; and AI, green stars) are compared with quantum calculations (black lines, QM). (a) Transmission probability for state 1 (T1) and state 2 (T2); (b) reflection probability for state 1 (R1) and state 2 (R2).

By introducing an augmented image, the interstate couplings are determined by local properties and can be well approximated even when the trajectories on inactive states move far away from the nonadiabatic coupling region so that MST-AI can provide a better description on nonadiabatic transitions especially for those bounced trajectories. Consequently the reflection probability is better reproduced, as shown in Fig. 1b. The MST-AI calculated $R_{1\leftarrow 1}$ is in excellent agreement with quantum results, but $R_{2\leftarrow 1}$ is a bit underestimated and appears roughly the same as $R_{1\leftarrow 1}$. The discrepancy from the exact results may be related to the classical trajectories used in the MST approach, and the prediction may be improved if multiconfigurations are used for each individual trajectory.[28,29,60] The oscillation of the reflection probabilities in the low momentum regime is due to the nuclear coherence overestimation similar to that in surface hopping although here it is much depressed. Note that MST-AI also predicts transmission probabilities nearly the same as the standard MST-QC approach (see Fig. 1a).

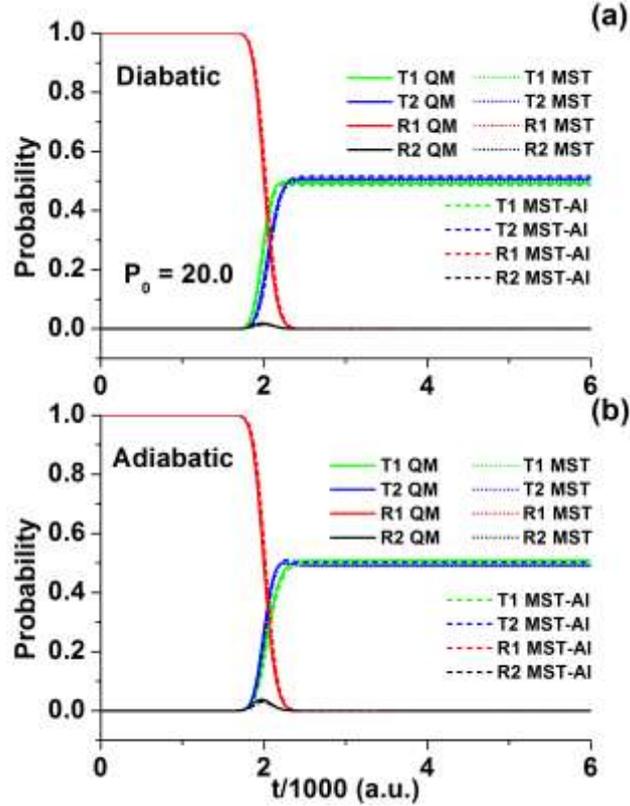

Figure 2. Time-dependent transition probability for the single avoided crossing problem. The results from the exact quantum DVR (solid), MST (full nuclear dynamics, dashed), and MST-AI (dotted) calculations are for the initial ground state transmission (T1, green), excited state transmission (T2, blue), ground state reflection (R1, red) and excited state reflection (R2, black). The initial momentum is $P_0 = 20.0$ a.u. (a) diabatic representation; (b) adiabatic representation. Note that here the final ground and excited states are switched in the adiabatic and diabatic representations.

To better evaluate the performance of the MST and its related approaches, we examine the time-dependent transition probabilities in both diabatic and adiabatic representations. Figure 2 and 3 display the results for the single and dual avoided crossing models, respectively, with an initial momentum $P_0 = 20.0$ a.u., in comparison with the exact quantum calculations. In these cases, agreement between the standard MST/MST-AI and quantum results in the whole time window examined are excellent, and the standard MST provides an essentially quantitative description on all channels, slightly better than what the MST-AI does.

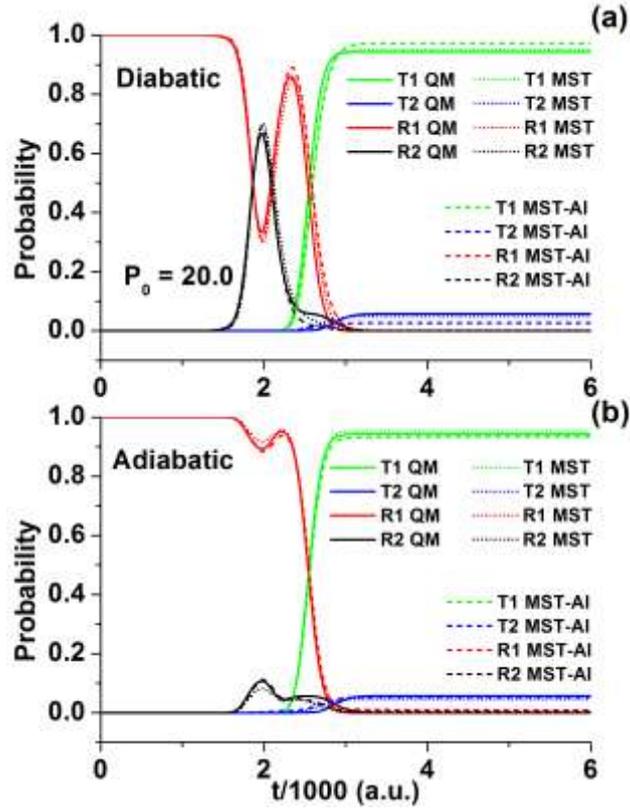

Figure 3. Same as Figure 2 except for the dual avoided crossing problem. The initial momentum is $P_0 =20.0$ a.u. (a) diabatic representation; (b) adiabatic representation.

For the extended coupling problem, only the adiabatic representation is considered, and the time-dependent transition probabilities with several different values of initial momentum, namely $P_0 = 3.0, 10.0, 20.0$, and $30.0$ a.u., are plotted in Figure 4 and . When the initial momentum is small, ex. $P_0 = 3.0$ a.u., the nonadiabatic effect is small. Most trajectories travel through the nonadiabatic coupling region and continue to move forward with only a small fraction of trajectories transit to the excited state and eventually bounce back (see Fig. 4a). In this case, both the standard MST-QC and MST-AI work reasonably well.

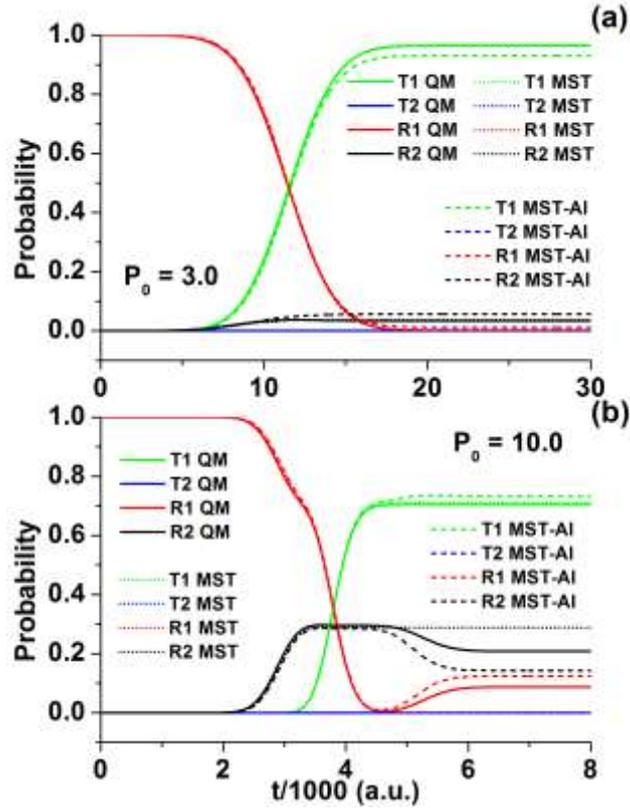

Figure 4. Time-dependent transition probability for the extended coupling problem. The results from the exact quantum DVR (solid), MST-QC (dashed), and MST-AI (dotted) calculations are for the ground state transmission (T1, green), excited state transmission (T2, blue), ground state reflection (R1, red) and excited state reflection (R2, black). The initial momentum is: (a) $P_0 =3.0$ a.u.; (b) $P_0 =10.0$ a.u..

As the initial momentum increases, ex. $P_0 = 10.0$ a.u., nonadiabatic transition becomes appreciable and the exact ground (excited) state reflection probability first decreases (increases) corresponding to the first nonadiabatic transition then increases (decreases) by some amount due to the second transition from the excited state to the ground state as the returned trajectory passes through the coupling region again (see Fig. 4b). Both the standard MST-QC and the MST-AI approaches produce nearly identical results in quantitative agreement with exact quantum calculations before the second transition, however the former fails to predict the second transition due to the insufficient estimation on the nonadiabatic coupling and all reflected trajectories stay on the excited state. By contrast, the MST-AI treatment is able to reproduce the reflection probabilities at least qualitatively correctly even though there is still room for improvement.

When the initial momentum approaches the threshold ($P_{th}$ = 28.3 a.u.), ex. $P_0$ = 20.0 a.u., between the first and second nonadiabatic transitions, a fraction of trajectories are able to climb up along the excited state potential and return before reaching the top of the barrier indicated by the dip and small peak in the excited state reflection and transmission probabilities around t = 2000 a.u. (see Fig. S4a). This feature and the first transition are well reproduced by both the standard MST and MST-AI treatments, and they produce different results mainly for the second transition, in a similar way as the case in Fig. 4b.

Once the momentum threshold is reached, transmissions on both states are allowed, and the reflection probability becomes small (see Fig. S4b). Both the MST-QC and MST-AI approaches perform pretty well again, presumably due to the small contribution from the multiple nonadiabatic transitions. For a high initial momentum such as $P_0$ = 30.0 a.u., the excited state transmission probability and the corresponding reflection probabilities reach a plateau before the equilibrium population is achieved at very long times. Therefore a large grid size of the system is required to produce corrected quantum results. This also exemplifies that better understandings of nonadiabatic processes could be achieved by real time dynamics rather than a single value of equilibrium or steady-state quantity such as transition probability.

Next we apply the MST approach to a two-state 2D quasi-Jahn-Teller problem. The system starts initially from the first adiabatic state in the positive R region of the R-r plane with an initial momentum in the negative R direction. As time involves, the time-dependent nonreactive probability (R>0) and reactive probability (R<0) are calculated. Figure 5 compares the results obtained from the MST and MST-AI approach along with the exact quantum DVR calculations. In diabatic representation (Fig. 5a), the standard MST method is capable of describing the passage of the system across the conical intersection qualitatively well, and the calculated probability show a dip in the nonreactive channel (a bump in the corresponding reactive channel) then reaching a similar plateau. The AI results also capture the transition dynamics however the amount is underestimated. By contrast, in adiabatic representation (Fig. 5b), the AI approach reproduces the quantum results almost perfectly. The standard MST tends to overestimate the transition probability. To do a further check, we change the initial condition of the system from a pure adiabatic state to a mixed state with an equal probability on both adiabatic states, and the results are shown in Figure 5c. Again the AI approach performs very well with the results are nearly identical to the exact calculations. The standard MST also shows excellent agreement with the quantum prediction for the early transition but underestimate the asymptotic probabilities. Our results here are comparable with the previous TDDVR work (see Figure 1a in ref 43).

Let us now consider a spin-boson problem in the nonadiabatic Marcus regime, which has been shown to be a strong test for the validity of detailed balance.[23] Figure 6 displays the time-dependent population of the initial state of the two-state spin-boson model for a nuclear mode coupled with a purely Ohmic bath. Results are calculated for a relatively long time (10 times longer than that in Ref 23) by Marcus theory, MST, MST-AI, SQC, and the coherence controlled (CC) nonadiabatic approach.[84] It is clear that both AI and CC approaches satisfy detailed balance quite well. Even the standard MST and SQC also seem to recover detailed balance at long times. It worth noting that in comparison the Marcus result (presumably accurate in this regime) the standard MST and SQC predict much slower dynamics, which deserves further investigations.

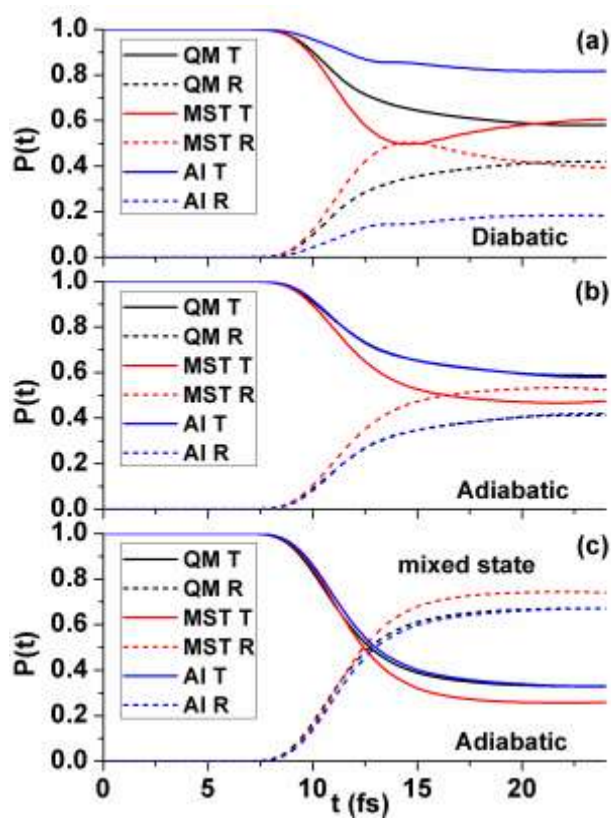

Figure 5. Time-dependent transition probability for the quasi-Jahn-Teller problem. The results from the exact quantum DVR (black), MST (red), and MST-AI (blue) calculations are for the nonreactive transition (T, solid lines), and reactive transition (R, dashed lines). The initial conditions are: $R_0 = 3.0$ a.u., $P_0 = -23.9$ a.u.. (a) diabatic representation; (b) adiabatic representation; (c) adiabatic representation for the initially mixed state.

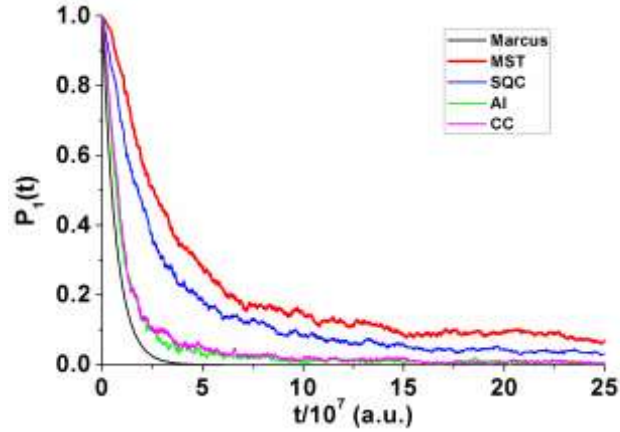

Figure 6. Time-dependent transition probability for the spin-boson problem. The results are obtained from the Marcus theory (black), MST (red), SQC (blue), AI (green), and coherent controlled (CC, pink) methods. 2000 bath modes, cutoff frequency $\omega_c = 400\Omega$, and dt =2 a.u.

As mentioned above, the MST and MST-AI approaches differ from the original and modified surface hopping methods in two aspects. First is how to describe coupled electronic-nuclear dyanmics. Although recently developed surface hopping methods introduce auxiliary variables[19] or spawning trajectories[20] to calculate a phenomenological decoherence rate, the nuclear dynamics is decoupled from electronic motions. By contrast in the standard MST approach, coupled nuclear-electronic dynamics is well described by either time dependent Schrödinger equation Eq. 3 or Heisenberg equation Eq. 9 with a nonlocal Hamiltonian matrix elements evaluated by multistate trajectories. Second is the way to handle nonadiabatic transitions. All MST approaches including the MST-AI which adopts a local Hamiltonian Eq. 14, determine the active state by quantizing the electronic state space using a window function proposed in the standard SQC approach[65] instead of the stochastic "hopping criterion" in surface hopping.[12] No momentum adjustment is required in the MST approaches since another individual trajectory or augmented image is always available on other inactive states. The energy flow between nuclear and electronic DOFs is inherently and self-consistently described by the coupled time dependent nuclear-electronic equations of motion, Eq. 3 or Eq. 9, so that there is no need to artificially adjust the momentum.

In the current implementation of the MST-QC and MST-AI approaches, we make a number of approximations. For example, in deriving the EOM for the coupled electronic-nuclear dynamics (Eq. 7), we leave out high order terms of $\hbar$ including the Bohmian quantum potential. Furthermore, nuclear dynamics is treated classically and phase information is lost. Consequently the quasi-classical and augmented image versions of MST may not describe well nuclear quantum effects such as tunneling or nuclear interference. In fact the standard MST formalism could represent nuclear dynamics in terms of wave packets[39] or semiclassical trajectories[73] like that in SC-IVR, which in principle could account for nuclear quantum effects. However, the computational cost would also be increasingly expensive especially for large systems, unless further approximations are used or efficient sampling schemes are applied. While for small systems, it would be better to just use the exact quantum methods based on grids or wave function. Nevertheless it would worth exploring in the future how to incorporate nuclear quantum effects into the MST formalism efficiently.

## V. Conclusions

In summary, the recently developed MST approach was derived in a general and rigorous framework starting from the time-dependent Schrödinger equation while incorporating both diabatic and adiabatic representations, as a complement to previous work. Ehrenfest or Born-Oppenheimer dynamics can be obtained from the MST formalism as the mean field or the single state limits, respectively. The resulting equations of motion are essentially in the same flavor with the ones in the SQC/MM methodology except for nuclear dynamics is in a multistate representation with one individual nuclear trajectory assigned to every involving state, which in principle can describe nuclear motions on individual state better than the mean field description in a variety of MM model based methods. By quantizing nuclear dynamics to a particular active state using a windowing technology proposed in the SQC/MM approach, the MST algorithm is numerically stable and does not suffer from the instability caused by the negative instant electronic population variables in the MM dynamics. On the other hand, the MST formalism is based on adiabatic nuclear basis (one trajectory for each state) therefore free of numerical instability caused by mixed nuclear states. Unlike the original or modified surface hopping methods, the MST approach incorporates electronic-nuclear coherence consistently and contains no ad hoc sudden state switch and the associated momentum adjustment or parameters to account for artificial decoherence. And the total energy of the MST representation is conserved and in connection with the measured total energy on the ensemble average level. When the active state trajectory ensemble is considered, the augmented image version of MST is achieved, which provides an illustrative understanding of the trajectory surface hopping treatment.

The MST approach is implemented to Tully's nonadiabatic problems and a quasi-Jahn-Teller problem involving conical intersection in both diabatic and adiabatic representations. The overall agreement between the MST results and exact quantum calculations is reasonably good and sometimes even quantitative in both representations. And it seems that MST-AI performs better in adiabatic representation. Despite a quasi-classical treatment, both MST-QC and AI describe nonadiabatic transitions quite accurately for Tully's models, especially for the first pass of the coupling region, which implies that for bounded condensed phase systems, the current version of MST approach could be accurate enough. For unbounded systems with multiple nonadiabatic transitions, such as the extended coupling problem considered in this work, the MST predicted transmission probabilities are in excellent agreement with the exact quantum results while the prediction on the reflection probability of the system with a substantial contribution from the second nonadiabatic transition is not satisfied, because for the multiple nonadiabatic transitions, individual trajectories move apart in the standard MST treatment. By contrast the MST-AI method is able to capture the second nonadiabatic transition and provide a reasonably good prediction on reflection probabilities. Moreover the MST-AI achieves extremely high accuracy in comparison with full quantum calculations for the quasi-Jahn-Teller problem in adiabatic representation although the agreement in the diabatic representation is also good. The MST-AI obeys detailed balance reasonably well while the standard MST satisfies detailed balance at least approximately.

In conclusion, the MST approach (especially its quasi-classical version and the derived AI version) features at least two notable advantages: consistent description of coupled electronic-nuclear dynamics and excellent numerical stability. Therefore it seems very promising to provide a practically efficient way for nonadiabatic dynamics simulations of realistic molecular systems, including ab inito molecular dynamics, which will be the future work.

AUTHOR INFORMATION

**Corresponding Author**

*Email: taogh@pkusz.edu.cn

**Notes**

The authors declare no competing financial interests.

ACKNOWLEDGMENT

We acknowledge the support from National Science Foundation of China (21673012 and 51471005) and Peking University Shenzhen Graduate School.

REFERENCES

1. W. H. Miller, Electronically Nonadiabatic Dynamics via Semiclassical Initial Value Methods. *J. Phys. Chem. A* **113**, 1405-1415 (2009).
2. J. C. Tully, Perspective: Nonadiabatic Dynamics Theory. *J. Chem. Phys.* **137**, 22A301 (2012).
3. T. Yonehara, K. Hanasaki, and K. Takatsuka, Fundamental Approaches to Nonadiabaticity: Toward a Chemical Theory beyond the Born Oppenheimer Paradigm. *Chem. Rev.* **112**, 499–542 (2012).
4. L. Wang, A. Akimov, and O. V. Prezhdo, Recent Progress in Surface Hopping: 2011−2015. *J. Phys. Chem. Lett.* **7**, 2100−2112 (2016).
5. P. Ehrenfest, Bemerkung Über Die Angenäherte Gültigkeit Der Klassischen Mechanik Innerhalb Der Quantenmechanik. *Z. Phys.* **45**, 455–57 (1927).
6. A. D. McLachlan, A variational solution of the time-dependent Schrodinger equation. *Mol. Phys.* **8**, 39-44 (1964).
7. G. D. Billing, On the Applicability of the Classical Trajectory Equations in Inelastic Scattering Theory. *Chem. Phys. Lett.* **30**, 391-393 (1975).
8. X. Li, J. C. Tully, H. B. Schlegel, and M. J. Frisch, *Ab Initio* Ehrenfest Dynamics. *J. Chem. Phys.* **123**, 084106 (2005).
9. A. W. Jasper, S. Nangia, C. Zhu, and D. G. Truhlar, Non-Born-Oppenheimer Molecular Dynamics. *Acc. Chem. Rev.* **39**, 101-108 (2006).
10. T. C. Berkelbach, T. E. Markland, and D. R. Reichman, Mixed Quantum-Classical Description of Excitation Energy Transfer in a Model Fenna−Matthews−Olsen Complex. *J. Chem. Phys.* **136**, 084104 (2012).
11. H. W. Kim, and Y. M. Rhee, Improving long time behavior of Poisson bracket mapping equation: A non-Hamiltonian approach. *J. Chem. Phys.* **140**, 184106 (2014).
12. J. C. Tully, Molecular Dynamics with Electronic Transitions. *J. Chem. Phys.* **93**, 1061-1071 (1990).
13. J. C. Tully, Mixed quantum classical dynamics. *Faraday Discuss.* **110**, 407-419 (1998).
14. M. D. Hack, and D. G. Truhlar, Nonadiabatic Trajectories at an Exhibition. *J. Phys. Chem. A* **104**, 7917-7926 (2000).
15. Y. Wu, and M. F. Herman, Nonadiabatic surface hopping Herman-Kluk semiclassical initial value representation method revisited: Applications to Tully's three model systems. *J. Chem. Phys.* **123**, 144106 (2005).
16. B. J. Schwartz, E. R. Bittner, O. V. Prezhdo, and P. J. Rossky, Quantum Decoherence and the Isotope Effect in Condensed Phase Nonadiabatic Molecular Dynamics Simulations. *J. Chem. Phys.* **104**, 5942-5955 (1994).

17. O. V. Prezhdo, and P. J. Rossky, Evaluation of Quantum Transition Rates from Quantum-Classical Molecular Dynamics Simulations. *J. Chem. Phys.* **107**, 5863−5878 (1997).
18. H. M. Jaeger, S. Fischer, and O. V. Prezhdo, Decoherence-induced surface hopping. *J. Chem. Phys.* **137**, 22A545 (2012).
19. J. E. Subotnik, and N. Shenvi, A new approach to decoherence and momentum rescaling in the surface hopping algorithm. *J. Chem. Phys.* **134**, 024105 (2011).
20. N. Shenvi, J. E. Subotnik, and W. Yang, Simultaneous-trajectory surface hopping: A parameter-free algorithm for implementing decoherence in nonadiabatic dynamics. *J. Chem. Phys.* **134**, 144102 (2011).
21. N. Shenvi, J. E. Subotnik, and W. Yang, Phase-corrected surface hopping: Correcting the phase evolution of the electronic wavefunction. *J. Chem. Phys.* **135**, 024101 (2011).
22. J. E. Subotnik, Fewest-Switches Surface Hopping and Decoherence in Multiple Dimensions. *J. Phys. Chem. A* **115**, 12083-12096 (2011).
23. B. R. Landry, M. J. Falk, and J. E. Subotnik, Communication: The correct interpretation of surface hopping trajectories: How to calculate electronic properties. *J. Chem. Phys.* **139**, 211101 (2013).
24. J. E. Subotnik, W. Ouyang, and B. R. Landry, Can we derive Tully's surface-hopping algorithm from the semiclassical quantum Liouville equation? Almost, but only with decoherence. *J. Chem. Phys.* **139**, 214107 (2013).
25. P. Shushkov, R. Li, and J. C. Tully, Ring polymer molecular dynamics with surface hopping. *J. Chem. Phys.* **137**, 22A549 (2012).
26. L. Wang, D. Trivedi, and O. V. Prezhdo, Global Flux Surface Hopping Approach for Mixed Quantum-Classical Dynamics. *J. Chem. Theory Comput.* **10**, 3598−3605 (2014).
27. L. Wang, A. E. Sifain, and O. V. Prezhdo, Fewest Switches Surface Hopping in Liouville Space. *J. Phys. Chem. Lett.* **6**, 3827−3833 (2015).
28. R. Baer, D. M. Charutz, R. Kosloff, and M. Baer, A study of conical intersection effects on scattering processes: The validity of adiabatic single-surface approximations within a quasi-Jahn–Teller model. *J. Chem. Phys.* **105**, 9141-9152 (1996).
29. M. Baer, S. H. Lin, A. Alijah, S. Adhikari, and G. D. Billing, Extended approximated Born-Oppenheimer equation. I. Theory. *Phys. Rev. A* **62**, 032506 (2000).
30. S. Adhikari, G. D. Billing, A. Alijah, S. H. Lin, and M. Baer, Extended approximated Born-Oppenheimer equation. II. Application. *Phys. Rev. A* **62**, 032507 (2000).
31. F. Agostini, A. Abedi, and E. K. U. Gross, Classical nuclear motion coupled to electronic non-adiabatic transitions. *J. Chem. Phys.* **141**, 214101 (2014).
32. R. Kapral, and G. Ciccotti, Mixed Quantum-Classical Dynamics. *J. Chem. Phys.* **110**, 8919−8929 (1999).

33. R. Kapral, Progress in the Theory of Mixed Quantum-Classical Dynamics. *Annu. Rev. Phys. Chem.* **57**, 129−157 (2006).
34. C. C. Martens and J.-Y. Fang, Semiclassical-limit molecular dynamics on multiple electronic surfaces. *J. Chem. Phys.* **106**, 4918−4930 (1997).
35. C. C. Martens, Communication: Fully coherent quantum state hopping. *J. Chem. Phys.* **143**, 141101 (2015).
36. Q. Shi and E. Geva, A derivation of the mixed quantum-classical Liouville equation from the influence functional formalism. *J. Chem. Phys.* **121**, 3393-3404 (2004).
37. M. H. Beck, A. Jackle, G. A. Worth, and H. D. Meyer, The Multiconfiguration Time-dependent Hartree (MCTDH) Method: a Highly Efficient Algorithm for Propagating Wavepackets. *Phys. Rep.* **324**, 1-105 (2000).
38. H. Wang, and M. Thoss, Multilayer Formulation of the Multiconfiguration Time-dependent Hartree Theory. *J. Chem. Phys.* **119**, 1289-1299 (2003).
39. Y. Wu, M. F. Herman, and V. S. Batista, Matching-pursuit/split-operator Fourier-transform simulations of nonadiabatic quantum dynamics. *J. Chem. Phys.* **122**, 114114 (2005).
40. T. J. Martinez, M. Ben-Nun, and R. D. Levine, Multi-Electronic-State Molecular Dynamics: A Wave Function Approach with Applications. *J. Phys. Chem.* **100**, 7884 (1996).
41. M. Ben-Nun, and T. J. Martinez, A multiple spawning approach to tunneling dynamics. *J. Chem. Phys.* **112**, 6113-6121 (2000).
42. P. Puzari, S. A. Deshpande, S. Adhikari, A quantum-classical treatment of non-adiabatic transitions. *Chem. Phys.* **300**, 305–323 (2004).
43. P. Puzari, B. Sarkar, S. Adhikari, Quantum-classical dynamics of scattering processes in adiabatic and diabatic representations. *J. Chem. Phys.* **121**, 707-721 (2004).
44. D. V. Shalashilin, Nonadiabatic dynamics with the help of multiconfigurational Ehrenfest method: Improved theory and fully quantum 24D simulation of pyrazine. *J. Chem. Phys.* **132**, 244111 (2010).
45. R. E. Wyatt, C. L. Lopreore, and G. Parlant, Electronic transitions with quantum trajectories. *J. Chem. Phys.* **114**, 5113-5116 (2001).
46. S. Garashchuk, V. A. Rassolov, and G. C. Schatz, Semiclassical nonadiabatic dynamics using a mixed wave-function representation. *J. Chem. Phys.* **123**, 174108 (2005).
47. B. F. E. Curchod, and I. Tavernelli, On trajectory-based nonadiabatic dynamics: Bohmian dynamics versus trajectory surface hopping. *J. Chem. Phys.* **138**, 184112 (2013).
48. J. C. Burant and J. C. Tully, Nonadiabatic dynamics via the classical limit Schrödinger equation. *J. Chem. Phys.* **112**, 6097-6103 (2000).
49. P. Pechukas, P. Time-Dependent Semiclassical Scattering Theory. II. Atomjc collisions. *Phys. Rev.* **181**, 174-184 (1969).
50. H. D. Meyer and W. H. Miller, A Classical Analog for Electronic Degrees of Freedom in Nonadiabatic Collision Processes. *J. Chem. Phys.* **70**, 3214-3223 (1979).

51. E. Neria and A. Nitzan, Semiclassical evaluation of nonadiabatic rates in condensed phases. *J. Chem. Phys.* **99**, 1109-1123 (1993).
52. G. Stock, A Semiclassical Self-Consistent-Field Approach to Dissipative Dynamics: the Spin-Boson Problem. *J. Chem. Phys.* **103**, 1561-1573 (1999).
53. G. Stock and M. Thoss, Semiclassical Description of Nonadiabatic Quantum Dynamics. *Phys. Rev. Lett.* **78**, 578-581 (1997).
54. M. Thoss and G. Stock, Mapping approach to the semiclassical description of nonadiabatic quantum dynamics. *Phys. Rev. A* **59**, 64–79 (1999).
55. A. A. Golosov and D. R. Reichman, Classical mapping approaches for nonadiabatic dynamics: Short time analysis. *J. Chem. Phys.* **114**, 1065–1074 (2001).
56. P. Huo and D. F. Coker, Partial Linearized Density Matrix Dynamics for Dissipative, Non-Adiabatic Quantum Evolution. *J. Chem. Phys.* **135**, 201101 (2011).
57. X. Sun and W. H. Miller, Semiclassical Initial Value Representation for Electronically Nonadiabatic Molecular Dynamics. *J. Chem. Phys.* **106**, 6346−6353 (1997).
58. X. Sun, H. Wang and W. H. Miller, Semiclassical Theory of Electronically Nonadiabatic Dynamics: Results of a Linearized Approximation to the Initial Value Representation. *J. Chem. Phys.* **109**, 7064–7074 (1998).
59. H. Wang, X. Song, D. Chandler and W. H. Miller, Semiclassical study of electronically nonadiabatic dynamics in the condensed-phase: Spin-boson problem with Debye spectral density. *J. Chem. Phys.* **110**, 4828–4840 (1999).
60. N. Ananth, C. Venkataraman, and W. H. Miller, W. H. Semiclassical Description of Electronically Nonadiabatic Dynamics via the Initial Value Representation. *J. Chem. Phys.* **127**, 084114 (2007).
61. J. O. Richardson and M. Thoss, Nonadiabatic Ring-Polymer Molecular Dynamics. *J. Chem. Phys.* **139**, 031102 (2013).
62. N. Ananth, Mapping Variable Ring Polymer Molecular Dynamics: A Path-Integral Based Method for Nonadiabatic Processes. *J. Chem. Phys.* **139**, 124102 (2013).
63. A. R. Menzeleev, F. Bell and T. F. III Miller, Kinetically Constrained Ring-Polymer Molecular Dynamics for Non-Adiabatic Chemical Reactions. *J. Chem. Phys.* **140**, 064103 (2014).
64. S. J. Cotton and W. H. Miller, Symmetrical Windowing for Quantum States in Quasi-Classical Trajectory Simulations. *J. Phys. Chem. A* **117**, 7190-7194 (2013).
65. S. J. Cotton and W. H. Miller, Symmetrical Windowing for Quantum States in Quasi-Classical Trajectory Simulations: Application to Electronically Non-Adiabatic Processes. *J. Chem. Phys.* **139**, 234112 (2013).
66. W. H. Miller and S. J. Cotton, Communication: Note on Detailed Balance in Symmetrical Quasi-Classical Models for Electronically Non-Adiabatic Dynamics. *J. Chem. Phys.* **142**, 131103 (2015).
67. W. H. Miller and S. J. Cotton, Communication: Wigner Functions in Action-angle Variables, Bohr-Sommerfeld Quantization, the Heisenberg Correspondence Principle, and a Symmetrical Quasiclassical Approach to the Full Electronic Density Matrix. *J. Chem. Phys.* **145**, 081102 (2016).

68. G. Tao, Electronically Non-adiabatic Dynamics in Singlet Fission: a Quasi-Classical Trajectory Simulation. *J. Phys. Chem. C* **118**, 17299-17305 (2014).
69. G. Tao, Bath Effect in Singlet Fission Dynamics. *J. Phys. Chem. C* **118**, 27258-27264 (2014).
70. G. Tao, Understanding Electronically Non-Adiabatic Relaxation Dynamics in Singlet Fission. *J. Chem. Theory Comput.* **11**, 28-36 (2015).
71. S. J. Cotton and W. H. Miller, The Symmetrical Quasi-Classical Model for Electronically Non-Adiabatic Processes Applied to Energy Transfer Dynamics in Site-Exciton Models of Light-Harvesting Complexes. *J. Chem. Theory Comput.*, **12**, 983–991 (2016).
72. W. H. Miller, The semiclassical initial value representation: A potentially practical way for adding quantum effects to classical molecular dynamics simulations. *J. Phys. Chem. A*. **105**, 2942-2955 (2001).
73. G. Tao, Electronically Non-Adiabatic Dynamics in Complex Molecular Systems: An Efficient and Accurate Semiclassical Solution. *J. Phys. Chem. A* **117**, 5821-5825 (2013).
74. G. Tao, A multi-state trajectory method for non-adiabatic dynamics simulations. *J. Chem. Phys.* **144**, 094108 (2016).
75. G. Tao, Nonequilibrium Electron-Coupled Lithium Ion Diffusion in $LiFePO_4$: Nonadiabatic Dynamics with Multistate Trajectory Approach. *J. Phys. Chem. C* **120**, 6938-6952 (2016).
76. D. Bohm, A Suggested Interpretation of the Quantum Theory in Terms of "Hidden" Variables. I. *Phys. Rev.* **85**, 166 (1952).
77. T. Yonehara and K. J. Takatsuka, Phase-space averaging and natural branching of nuclear paths for nonadiabatic electron wavepacket dynamics. *J. Chem. Phys.* **129**, 134109 (2008).
78. N. Makri and W. H. Miller, Time-dependent self-consistent field (TDSCF) approximation for a reaction coordinate coupled to a harmonic bath: Single and multiple configuration treatments. *J. Chem. Phys.* **87**, 5781– 5787 (1987).
79. G. Tao, Supplemental Material.
80. G. Tao and W. H. Miller, Abstracts of Papers, 242[nd] National Meeting of the American Chemical Society, Denver, CO, Aug 28-Sep 1, 2011; American Chemical Society: Washington, DC, 2011; PHYS 371.
81. M. Born and E. Oppenheimer, Zur Quantentheorie der Molekeln. *Ann. Phys.* **84**, 457-484 (1927).
82. W. H. Miller, Perspective: Quantum or Classical Coherence? *J. Chem. Phys.* **136**, 210901 (2012).
83. S. Bai, K. Song, and Q. Shi, Effects of Different Quantum Coherence on the Pump−Probe Polarization Anisotropy of Photosynthetic Light-Harvesting Complexes: A Computational Study. *J. Phys. Chem. Lett.* **6**, 1954−1960 (2015).
84. G. Tao, Coherence Controlled Non-Adiabatic Dynamics via State Space Decomposition: a Consistent Way to Incorporate Ehrenfest and Born-Oppenheimer-like Treatments on Nuclear Motion, *J. Phys. Chem. Lett.* **7**, 4335-4339 (2016).

85. D. T. Colbert and W. H. Miller, A Novel Discrete Variable Representation for Quantum Mechanical Reactive Scattering via the *S*-Matrix Kohn Method. *J. Chem. Phys.* **96**, 1982– 1991 (1992).

Table of Contents

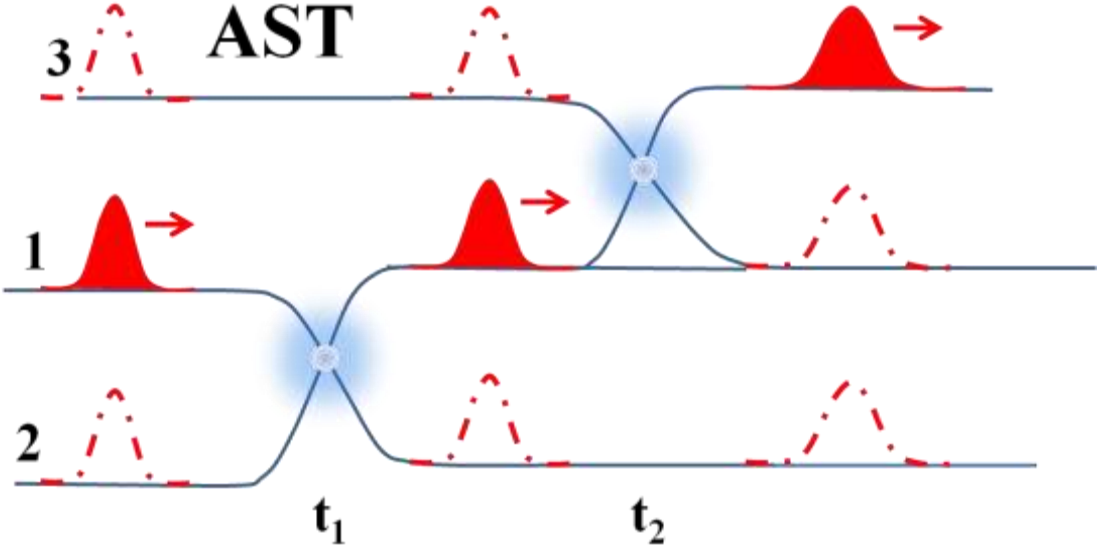